\documentclass[11pt]{article}

\usepackage{euscript}
\usepackage{amssymb}
\usepackage{amsfonts}
\usepackage{amsbsy}
\usepackage{epsfig}
\usepackage{amsthm}
\usepackage{amscd}
\usepackage{amstext}
\usepackage{verbatim}
\usepackage{amsmath}
\usepackage{cancel}
\usepackage{capt-of}
\usepackage{empheq}
\usepackage{graphicx}
\usepackage{caption}
\usepackage{subcaption}
\usepackage[cal=boondoxo]{mathalfa}

\usepackage[pdftex]{hyperref}

\def\ben{\begin{equation}}
\def\een{\end{equation}}

\let\a=\alpha

\def\be{\begin{equation}}
\def\ee{\end{equation}}
\def\beq{\begin{equation}}
\def\eeq{\end{equation}}
\def\ba{\begin{array}}
\def\ea{\end{array}}

\def\dalemb#1#2{{\vbox{\hrule height .#2pt
       \hbox{\vrule width.#2pt height#1pt \kern#1pt
               \vrule width.#2pt}
       \hrule height.#2pt}}}

\newcommand{\bea}{\begin{eqnarray}}
\newcommand{\eea}{\end{eqnarray}}

\def\R{{{\Bbb R}}}

\begin{document}

\begin{center}

{ \Large {\bf
Cold planar horizons are floppy
}
}

\vspace{1cm}

Sean A. Hartnoll$^{1}$ and Jorge E. Santos$^{1,2}$

\vspace{1cm}

{\small
$^{1}${\it Department of Physics, Stanford University, \\
Stanford, CA 94305-4060, USA }}

\vspace{0.5cm}

{\small
$^{2}${\it Department of Applied Mathematics and Theoretical Physics, \\
University of Cambridge, Wilberforce Road, \\
Cambridge CB3 0WA, UK}}

\vspace{1.6cm}

\end{center}

\begin{abstract}

Extremal planar black holes of four dimensional Einstein-Maxwell theory with a negative cosmological constant have an 
AdS$_2 \times \R^2$ near horizon geometry. We show that this near horizon geometry admits a deformation to a two parameter family of extremal geometries with inhomogeneous, spatially periodic horizons. At a linear level, static inhomogeneous perturbations of AdS$_2 \times \R^2$ decay towards the horizon and thus appear irrelevant under the holographic RG flow. However we have found numerically that
nonlinear effects lead to inhomogeneous near horizon geometries. A consequence of these observations is that an arbitrarily small periodic deformation of the boundary theory at nonzero charge density does not flow to AdS$_2 \times \R^2$ in the IR, but rather to an inhomogeneous horizon. These results shed light on existing numerical studies of low temperature periodically modulated black holes and also offer a new mechanism for holographic metal-insulator crossovers or transitions.

\end{abstract}

\pagebreak
\setcounter{page}{1}

\section{Introduction}

The near horizon geometries of black holes in Anti-de Sitter spacetime describe the low energy dissipative dynamics of strongly interacting quantum field theories. Of particular recent interest have been the near horizon geometries of extremal planar black holes, as these dually capture the dissipative dynamics of novel phases of zero temperature quantum matter \cite{Hartnoll:2011fn}. Such phases are of possible relevance to unconventional strongly interacting condensed matter systems.

An intriguing and ubiquitous character appearing in the investigations of extremal black holes as quantum matter is the near horizon AdS$_2 \times \R^2$ solution of Einstein-Maxwell theory with a negative cosmological constant \cite{Romans:1991nq, Chamblin:1999tk, Hartnoll:2011fn}. This solution has an exotic $z=\infty$ scaling symmetry under which time scales but space does not. That is, roughly speaking, the horizon supports low energy excitations with arbitrarily large momentum. This fact has two known important and phenomenologically interesting consequences. Firstly, these horizons can efficiently absorb the low energy excitations of a Fermi sea that carry a finite momentum \cite{Faulkner:2009wj} (following \cite{Lee:2008xf, Liu:2009dm, Cubrovic:2009ye}). Secondly, electric currents along the horizon can be efficiently degraded by finite wavevector lattice scattering \cite{Hartnoll:2012rj, Horowitz:2012ky}.

In this paper we uncover a further dramatic effect of nonzero wavevector dynamics on AdS$_2 \times \R^2$ horizons. We show that AdS$_2 \times \R^2$ admits a fully nonlinear deformation, with a tunable amplitude, in which translation invariance along the horizon is broken by a periodic function. Thus AdS$_2 \times \R^2$ itself appears as the near horizon geometry of one limit of a family of solutions with a finite inhomogeneity of the horizon. The existence of these solutions is rather nontrivial: at a linearized level, static nonzero wavevector perturbations of AdS$_2 \times \R^2$ show a power law decay towards the horizon \cite{Edalati:2010hk, Edalati:2010pn}. However, through fully nonlinear numerical studies, we will see that these modes source inhomogeneous terms that remain finite at the horizon. That is, AdS$_2 \times \R^2$ is not linearization stable and the irrelevant finite momentum couplings are dangerous.

The existence of nonlinear inhomogeneous deformations of the AdS$_2 \times \R^2$ near horizon geometry is an RG flow instability in the following sense. Start with an asymptotically AdS$_4$ spacetime and deform it by, for instance, a periodic source for the boundary electrostatic potential:
\begin{equation}
A^{(0)}_t = \bar{\mu} + \a \cos \left( k_L \, x \right) \,.
\label{eq:latticesource}
\end{equation}
The exact form of the asymptotic source is not too important, the essential physics is contained within the IR near horizon geometry.
Without the lattice deformation, the full spacetime is the extremal planar Reissner-Nordstr\"om-AdS (RN) solution and in particular the near horizon geometry is AdS$_2 \times \R^2$. Upon adding the periodic deformation (\ref{eq:latticesource}), a first expectation is that the near horizon geometry would likely remain AdS$_2 \times \R^2$, at the very least for small lattice amplitudes. This is because in Einstein-Maxwell theory all finite momentum deformations of AdS$_2 \times \R^2$ are irrelevant to linear order in perturbation theory \cite{Edalati:2010hk, Edalati:2010pn}. If this expectation were true, then AdS$_2 \times \R^2$ would be an IR attractive fixed point for the finite charge density theory, stable under UV deformations by a periodic potential. However, the results of this paper show that this is not the case. Instead, at least for some range of lattice wavevector $k_L$ and lattice amplitude $\a$, the IR fixed point moves continuously away from AdS$_2 \times \R^2$ as the lattice is turned on and becomes one of the inhomogeneous geometries we have mentioned. That is, we have a line of inhomogeneous IR fixed point geometries, and AdS$_2 \times \R^2$ itself is only reached if there is no lattice in the UV. These facts are visualized in the following figure \ref{fig:sketch}.

\begin{figure}[h]
\centering
\includegraphics[height = 0.32\textheight]{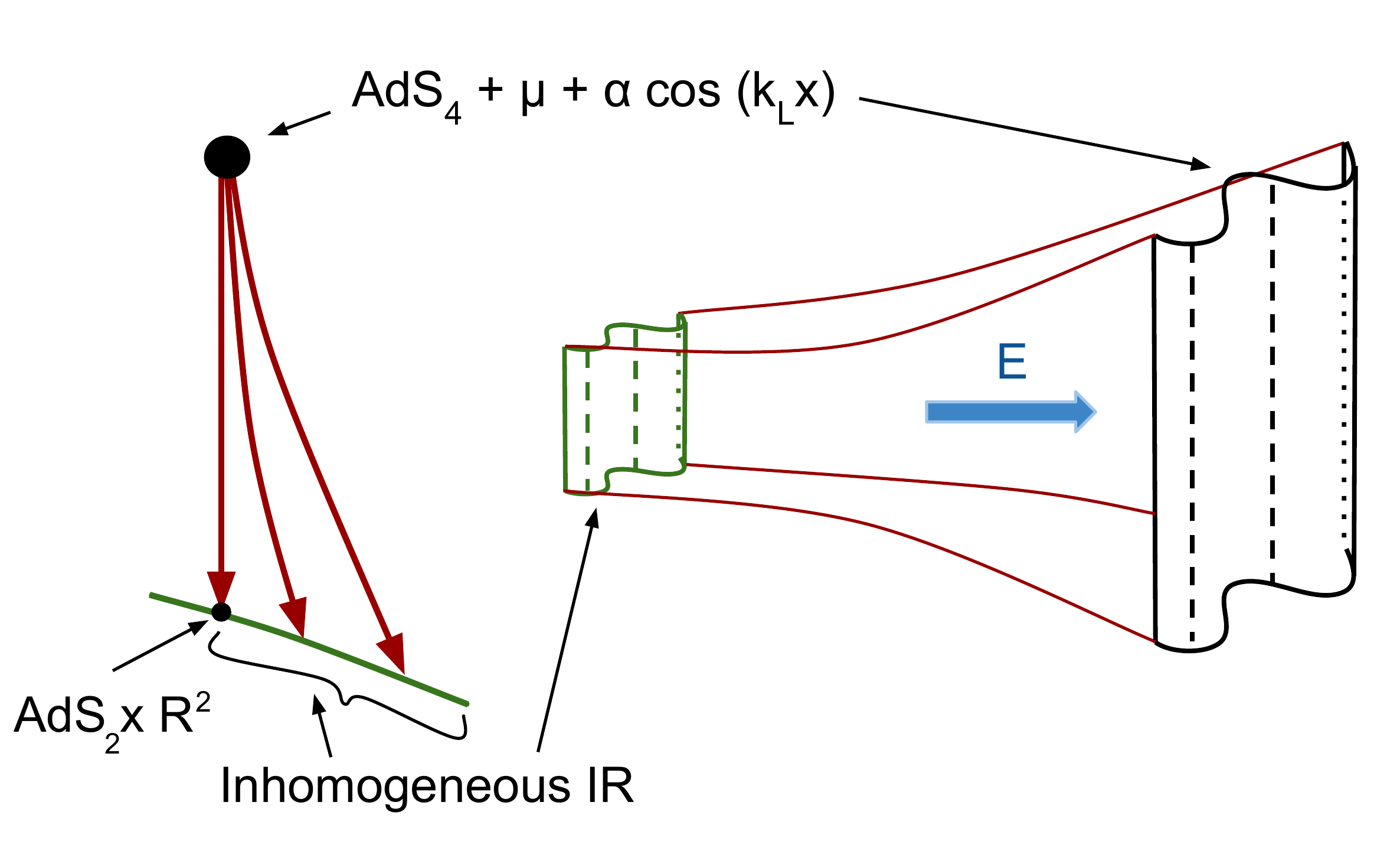}
\caption{\label{fig:sketch} In Einstein-Maxwell theory, AdS$_4$ deformed by a net chemical potential and a periodic source flows in the IR to a line of fixed points described by inhomogeneous extremal horizons. AdS$_2 \times \R^2$ only arises as the IR if there is no periodic source.}
\end{figure}

In section \ref{sec:numerics} of this paper we construct numerical zero temperature solutions in which a UV periodic potential results in an inhomogeneous horizon. In section \ref{sec:ads4} we show that at zero charge density (that is, $\bar{\mu} = 0$ in (\ref{eq:latticesource})) the corresponding AdS$_4$ spacetime is RG stable against the lattice perturbation, so that the IR geometry remains AdS$_4$ in this case. In the concluding discussion we outline the condensed matter phenomenology likely to follow from the new IR geometries.

Our new near horizon geometries retain the finite entropy density of AdS$_2 \times \R^2$. While this zero temperature entropy density is widely considered a pathology, the remarkable phenomenology of $z=\infty$ scaling geometries -- perhaps the richest and most novel output from holographic studies of quantum matter -- might advise against throwing out the baby with the bathwater. On the one hand, going slightly beyond Einstein-Maxwell theory allows for conformally AdS$_2 \times \R^2$ horizons that retain the $z=\infty$ phenomenology without the entropy density \cite{Gouteraux:2011ce, Anantua:2012nj, Hartnoll:2012wm, Gubser:2009qt}. On the other hand, extensive ground state entropies have also arisen in large $N$ theories of spin liquid phases \cite{spinglass}. Similarities between this last system and AdS$_2 \times \R^2$, pointed out in \cite{Sachdev:2010um}, may suggest that the ground state entropy density is trying to tell us something about strongly interacting densities of charge carriers, or at least that AdS$_2 \times \R^2$ will ultimately find its place within a larger family of large $N$ phases of matter. This last possibility offers to shed microscopic light on extremal non-supersymmetric black holes.

\section{\label{vorspiel}Setup}

In this paper we study certain black hole solutions to the Einstein-Maxwell-AdS theory
\begin{equation}
S= \frac{1}{16 \pi G_N}\int d^4 x\,\sqrt{-g}\left[R+\frac{6}{L^2}-\frac{1}{2}F_{ab}F^{ab}\right],
\label{eq:action}
\end{equation}
where  $L$ is the AdS length scale, $F= \mathrm{d} A$ and $G_N$ is Newton's constant. The equations of motion are
\begin{equation}
\begin{gathered}
R_{ab}+\frac{3}{L^2} g_{ab}-\left(F_{ac}F_{b\phantom{c}}^{\phantom{b}c}-\frac{g_{ab}}{4}F_{cd}F^{cd}\right) = 0 \,, \,
\\
\nabla_a F^{ab}  = 0\,.
\end{gathered}
\label{eqs:motion}
\end{equation}

We will primarily be interested in planar black hole solutions to the equations of motion (\ref{eqs:motion}) at zero and low temperatures. The basic and well known solution in this class is the extremal planar RN black hole, whose line element and gauge field can be written as
\begin{equation}
\begin{gathered}
\mathrm{d}s^2 = \frac{L^2}{y^2}\left[-(1+2y+3y^2)(1-y)^2\mathrm{d}t^2+\frac{\mathrm{d}y^2}{(1+2y+3y^2)(1-y)^2}+\mathrm{d}x^2+\mathrm{d}w^2\right] \,, \\
A = L\,\sqrt{6}\,(1-y)\mathrm{d}t\,.
\end{gathered}\label{eq:RN}
\end{equation}
The asymptotically AdS$_4$ conformal boundary is at $y \to 0$. The near horizon geometry is obtained by writing 
$t = \tau/\varepsilon$, $y = 1-\varepsilon \rho/6$ and taking the limit $\varepsilon\to0$. The resulting metric is recognized as being that of AdS$_2\times \R^2$, with an AdS$_2$ radius of $L_2 \equiv L/\sqrt{6}$:
\begin{equation}
\begin{gathered}
\mathrm{d}s^2 = L^2\left[\frac{1}{6}\left(-\rho^2\mathrm{d}\tau^2+\frac{\mathrm{d}\rho^2}{\rho^2}\right)+\mathrm{d}x^2+\mathrm{d}w^2\right] \,, \\
A = \frac{L\,\rho}{\sqrt{6}}\mathrm{d}\tau\,.
\end{gathered}
\label{eq:ads2r2}
\end{equation}

\section{\label{sec:numerics}Inhomogeneous AdS$_2$ horizons}

In this section we find the extremal solution numerically for several values of 
\be
k_0\equiv \frac{k_L}{\bar{\mu}} \,, \qquad A_0\equiv \frac{\alpha}{\bar{\mu}} \,. \label{eq:ratios}
\ee
That is, we will find full asymptotically AdS$_4$ solutions deformed by the UV source (\ref{eq:latticesource}).
As we will see, this is a challenging numerical calculation. In order to make sure our results are trustworthy, we also construct the full geometry at finite, but very small, temperature where more standard methods are applicable \cite{Horowitz:2012ky,Horowitz:2012gs}. Let us focus first on the zero temperature solution.

We are interested in black hole solutions with a timelike static Killing vector field $\partial_t$, and a spacelike Killing vector field $\partial_w$. Furthermore, we want the black hole to have a regular degenerate Killing horizon, \emph{i.e.} zero temperature. This restricts the most general line element to take the following form:
\begin{equation}
\begin{gathered}
\mathrm{d}s^2 = \frac{L^2}{y^2}\left[-(1-y)^2 \,G(y)A \mathrm{d}t^2+\frac{B}{(1-y)^2\,G(y)}(\mathrm{d}y+F\, \mathrm{d}x)^2+S_1 \mathrm{d}x^2+S_2\mathrm{d}w^2\right]
\\
A = L\,\sqrt{6}\,(1-y)\,P\,\mathrm{d}t\,.
\end{gathered}
\end{equation}
where $G(y)=1+2y+3y^2$. Here, $A$, $B$, $F$, $S_1$, $S_2$ and $P$ are functions of both $x$ and $y$ to be determined in what follows. Furthermore, $y=0$ denotes the conformal boundary and $y=1$ the location of the degenerate horizon. The factors of $(1-y)^2$ ensure that the line element above has a degenerate Killing horizon. Finally, $G(y)$ and the factor of $\sqrt{6}$ in the definition of $P$ were chosen such that when $A=B=S_1=S_2=P=F+1=1$, the line element above reduces to that of the RN black hole (\ref{eq:RN}).

Throughout the paper we will be interested in black hole solutions whose conformal boundary metric, located at $y=0$, is
\begin{equation}
\mathrm{d}s^2_{\partial} = -\mathrm{d}t^2+\mathrm{d}x^2+\mathrm{d}w^2\,.
\end{equation}
Furthermore, the function $P$ is such that at $y=0$, the dual electrostatic potential is
\begin{equation}
A_t^{(0)} = \bar{\mu} +\alpha \cos \left( k_L \, x \right)\,.
\end{equation}
Since the underlying UV microscopic theory is scale invariant, all physical observables will only depend on the dimensionless ratios (\ref{eq:ratios}). Our zero temperature parameter space is thus two dimensional.

Before detailing the numerical scheme we used to integrate Einstein's equations, we should discuss regularity at the extremal horizon, \emph{i.e.} $y=1$. Ingoing Eddington-Finkelstein coordinates take the form:
\begin{equation}
\begin{gathered}
t = v+\frac{1}{6}\frac{1}{1-y}-\frac{2}{9}\log(1-y)+\mathcal{O}(1) \,,
\\
x = X+\mathcal{O}((1-y))\,,
\end{gathered}
\end{equation}
where higher order terms are determined demanding $g_{yy}=0$ and $g_{Xy}=0$. Regularity at $y=1$ is then seen to require $A(x,1)=B(x,1)$ and $F(x,1)=0$, together with all remaining metric functions being finite at $y=1$. At no point here have we used Einstein's equations. This minimum set of boundary conditions had better be allowed by our integration scheme if we are to find a regular black hole solution, given our UV boundary conditions.

Our approach to solving Einstein's equations is similar to the one used in \cite{Horowitz:2012ky}, which was first introduced in \cite{Headrick:2009pv} and studied in great detail in \cite{Figueras:2011va}. We shall only review the main differences and difficulties. It is straightforward to deduce that there is non-analytic behavior close to the conformal boundary. This appears for instance in the cross term $F$ at order $y^2 \log y$. In order to deal with this, we introduce a finite difference patch close to the boundary. Perhaps more worrying, there is also non-analytic behavior at the horizon: if there were none, the near horizon geometry would have to reduce to the one presented in Section 6.2 of \cite{Kunduri:2013gce}. Furthermore, if the near horizon geometry were AdS$_2 \times \R^2$ exactly, we know from perturbative results \cite{Edalati:2010hk, Edalati:2010pn} that the approach to the horizon would involve terms of the form $(1-y)^{\nu}$, where, in general, $\nu$ is an irrational number. Our code must thus deal with these more general situations. We again introduce a finite difference patch, now close to the horizon, to ensure we can capture such non-analytic behavior. This type of singular behavior is more dangerous than the logarithmic behavior close to the conformal boundary. In particular, if no finite difference patch is used, for sufficiently small $\nu_-$ the spectral approximation ceases to converge at the horizon and starts diverging exponentially with increasing number of grid points! We also monitor the gradients as the Newton-Raphson method relaxes down to equilibrium. Whenever these are large, we double the number of grid points in both of the finite difference patches, \emph{i.e.} we use adaptive mesh refinement. We have also explicitly checked that if we \emph{only} use spectral methods, with none of the above improvements, the Newton-Raphson method \emph{does} converge, but to a solution that is not smooth and does not satisfy Einstein's equations, with the error being larger at the location of the horizon. Note that if both patches are sufficiently small, we should recover the exponential converge of the spectral method, for sufficiently large number of grid points $N$. In all of our simulations, the spectral patch contains no fewer than $70\times70$ grid points in both $y$ and $x$ direction. Along the $y$ direction we use the Gauss-Lobatto-Chebyshev grid and in $x$ we use the Fourier nodes. Finally, we vary the position of our patches (always keeping both substantially smaller than the spectral patch), and check that our results do not change within some specified precision.

We are especially interested in measuring deviations of the IR geometry from AdS$_2\times\R^2$. In particular, we would like to claim that the new near horizon geometry breaks translational invariance along the inhomogeneous field theory direction $x$. In order to quantify inhomogeneity of the horizon, we introduce the following quantity: $\mathcal{W} \equiv (\partial_w)^a (\partial_w)^b g_{ab}$ evaluated at the degenerate Killing horizon, \emph{i.e.} the norm squared of $\partial_w$ evaluated at the horizon. For AdS$_2\times\R^2$, the norm of $\mathcal{W}$ is constant. To see a sharp deviation from AdS$_2\times\R^2$ in the IR we decided to plot
\begin{equation}
\varpi\equiv\frac{\mathcal{W}_{\max}}{\mathcal{W}_{\min}}-1\,.
\label{eq:varpi}
\end{equation}
Any deviation of $\varpi$ from $0$ indicates that the near horizon geometry is \emph{not} that of an extreme RN black hole and furthermore is not translation invariant.

\begin{figure}[h]
\centering
	\begin{subfigure}{0.4\textwidth}
		\includegraphics[height = 0.26\textheight]{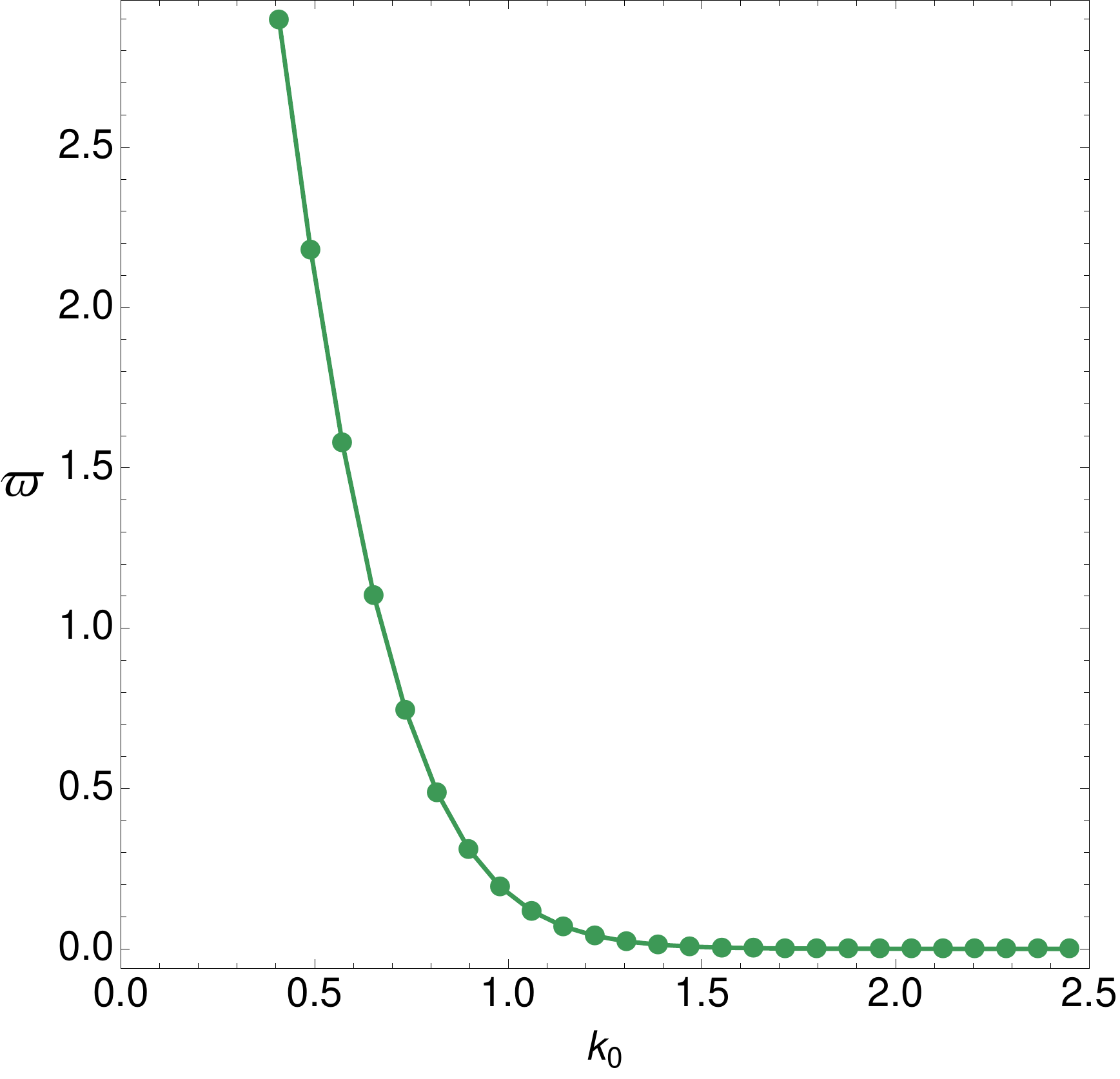}
		\caption{\label{fig:1a}Plot of $\varpi$, as a function of $k_0$, for fixed $A_0 = 1/2$.}
	\end{subfigure}
	\hspace{1 cm}
	\begin{subfigure}{0.4\textwidth}
		\includegraphics[height = 0.26\textheight]{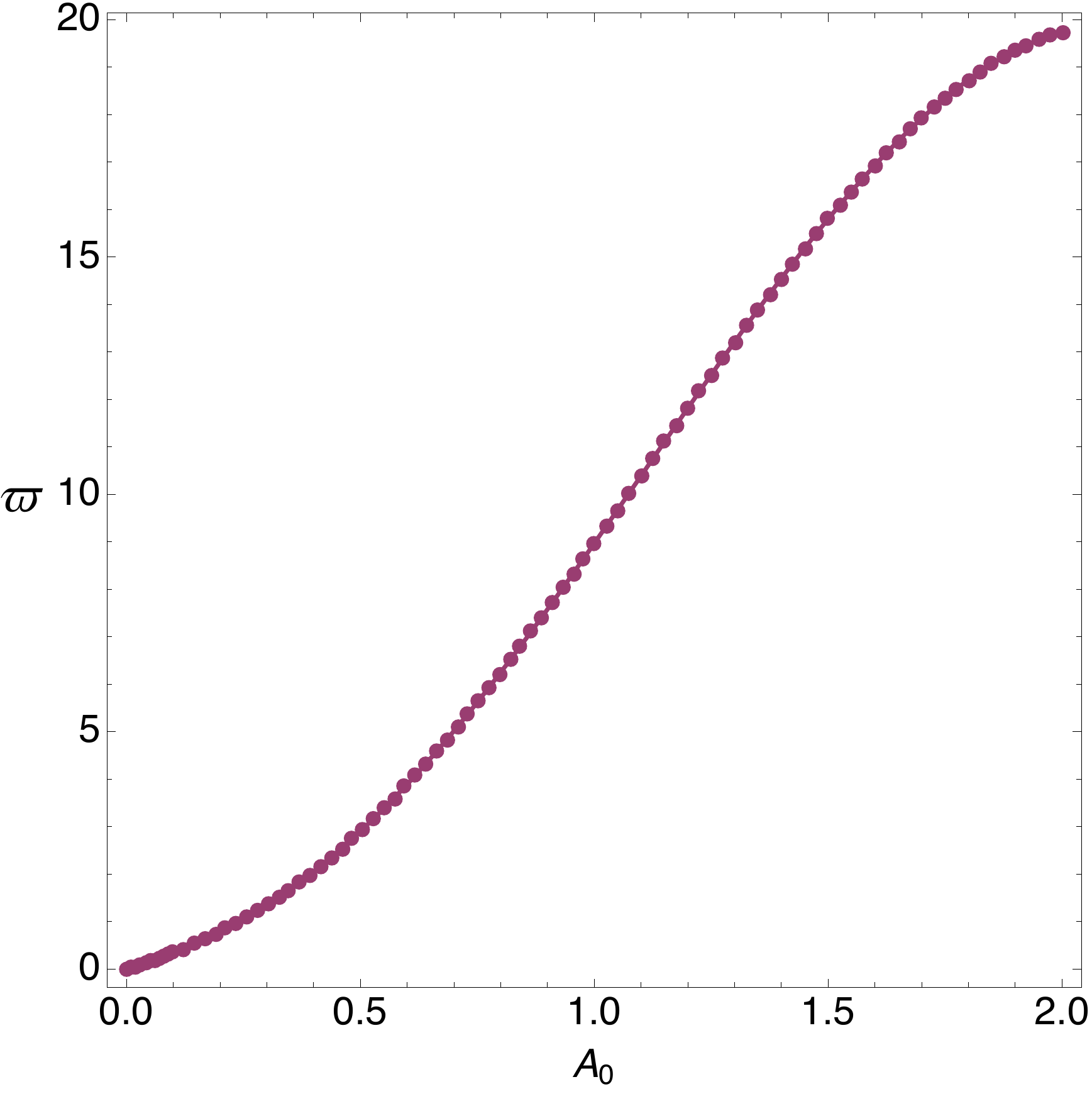}
		\caption{\label{fig:1b}Plot of $\varpi$, as a function of $A_0$, for fixed $k_0 = 1/\sqrt{6}$.}
	\end{subfigure}
	\caption{\label{fig:1}Plots of $\varpi$ -- the ratio (\ref{eq:varpi}) of the maximal and minimal values of $g_{ww}$ on the horizon (minus one) and therefore a measure of the inhomogeneity of the horizon -- as a function of the dimensionless quantities $k_0$ or $A_0$.}
\end{figure}

Fig.~\ref{fig:1a} shows $\varpi$ as a function of $k_0$, keeping $A_0=1/2$. We have repeated this calculation for several values of $A_0$, up to $2$, and we see no qualitative difference. The quantity $\varpi$ is clearly nonzero over a range of values of $k_0$, proving the existence of a new inhomogeneous near horizon geometry in these cases! We have not ruled out the existence of a critical $k_0$ beyond which the horizon returns to
AdS$_2\times\R^2$.
The increase of $\varpi$ at small $k_0$ is due to the fact that we have taken this limit with $A_0$ fixed. To recover the homogeneous case (with no lattice deformation), one must take $k_0$ to zero with $A_0/k_0$ fixed.

We have also studied $\varpi$ as a function of the amplitude $A_0$, now with $k_0$ kept fixed: this is depicted in Fig.~\ref{fig:1b}. This plot also confirms that a nonlinear RG instability indeed exists, and that it persists all the way down to \emph{any} value of $A_0>0$, including arbitrarily small lattice amplitudes. At small enough $A_0$, $\varpi$ is a linear function of $A_0$, the precise coefficient of which depends on $k_0$. In appendix \ref{sec:AA} we show the same plot with a larger value of $k_0$.

\begin{figure}[h]
\centering
\includegraphics[height = 0.22\textheight]{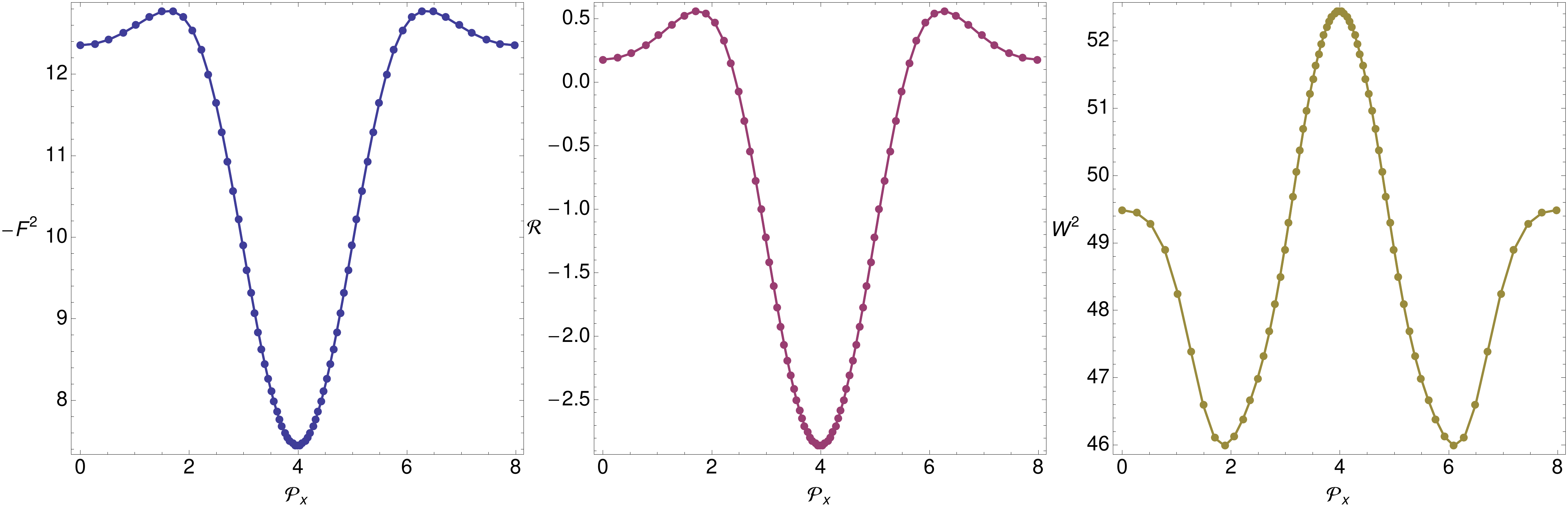}
\caption{\label{fig:2}Plots of the electric field squared on the horizon $F^2$, the induced scalar curvature of the horizon $\mathcal{R}$ and the Weyl tensor squared on the horizon $W^2$, as a function of the proper distance $\mathcal{P}_x$ along the horizon. All plots have $A_0=2$ and $k_0 = 1/\sqrt{6}$.}
\end{figure}

We have also extracted geometric invariants of the horizon: the electric field squared at the horizon $F^2$, the Ricci scalar of the induced horizon geometry $\mathcal{R}$ and the Weyl tensor squared of the bulk spacetime geometry $W^2 \equiv C^{abcd}C_{abcd}$ evaluated on the horizon. For the near horizon of planar RN black holes, \emph{i.e.} for AdS$_2\times\R^2$, these are, respectively, $-6$, $0$ and $48$, in AdS length units. In Fig.~\ref{fig:2} we show how these quantities look when evaluated in the new IR geometry, for $A_0=2$ and $k_0 = 1/\sqrt{6}$. The results once more demonstrate that the near horizon geometry is not AdS$_2\times \R^2$ but rather some wiggly version thereof. A natural candidate for such a near horizon geometry was presented in Section 6.2 of \cite{Kunduri:2013gce}, where the most general near horizon  compatible with a $C^2$ extremal geometry is found. That geometry is essentially a double Wick rotation of the charged four-dimensional hyperbolic black hole \cite{Lemos:1994xp}. We have attempted to match the invariants of Fig.~\ref{fig:2} with those from that near horizon geometry, with no success. We believe they are not the same geometry. This would seem to indicate that the new extremal geometries we have found are not $C^2$, enabling them to evade the general classification of \cite{Kunduri:2013gce}.
Our results motivate a revisiting of the classification theorems of extremal horizons, in particular with weakened analyticity assumptions near the horizon.

\begin{figure}[h]
\centering
\includegraphics[height = 0.29\textheight]{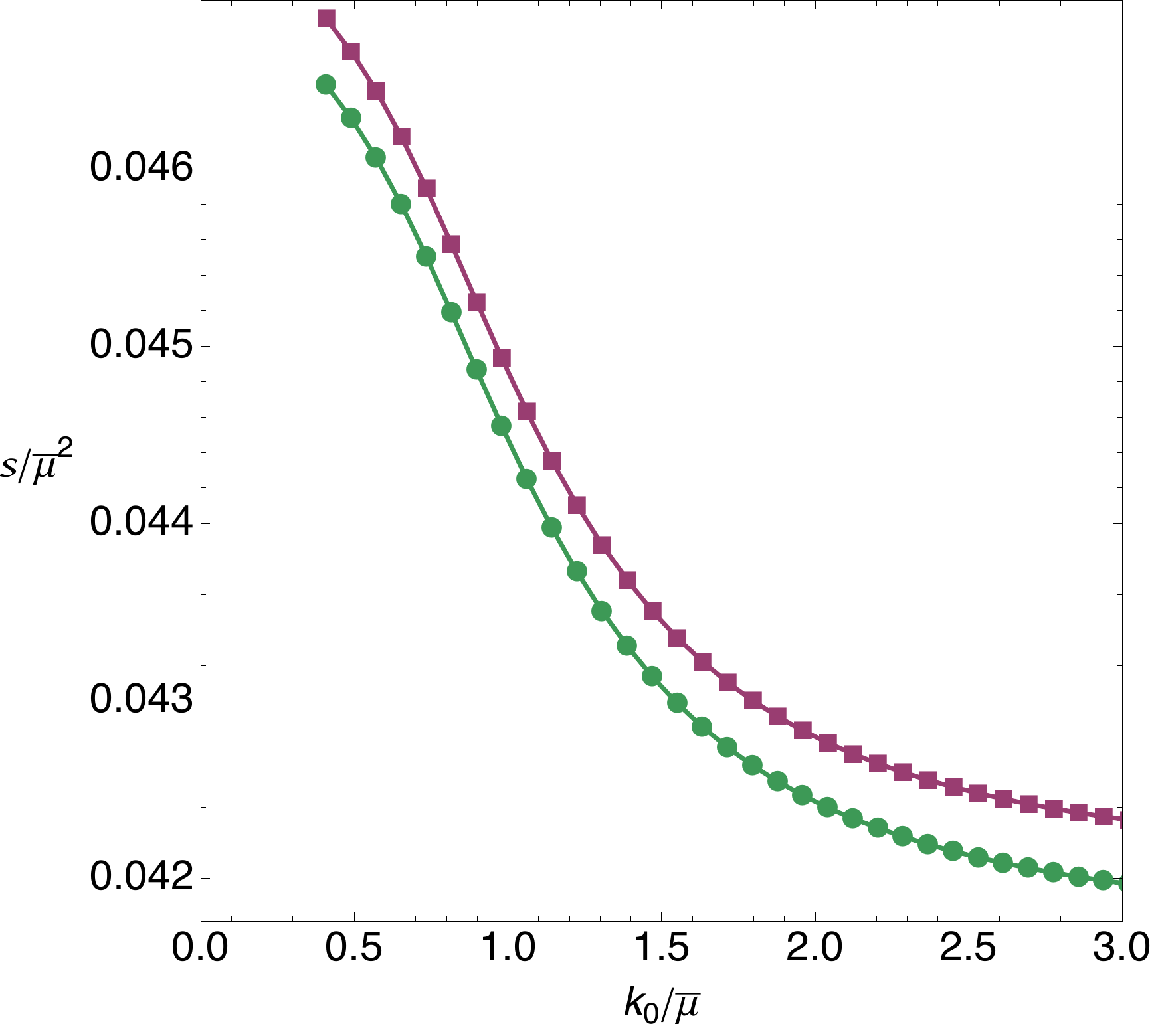}
\caption{\label{fig:4}Comparing the averaged entropy density $s$ at $T = 0$ (disks) and $T/\bar{\mu}\approx0.002$ (squares). In both cases, $A_0 = 1/2$.}
\end{figure}

We have cross checked all of our zero temperature calculations by solving the system at finite but very small temperature $T$ and checking that all the curves presented in this paper are readily approached as the temperature is lowered. The finite temperature calculations are very similar to those done in \cite{Horowitz:2012gs}, and we shall only present one illustrative final result. In Fig.~\ref{fig:4} we plot the average entropy density $s$ for $T=0$ (disks) and for $T/\bar{\mu} \approx 0.002$ (squares). The agreement between the two curves is reassuring.
In addition, Fig.~\ref{fig:4} allows us to emphasize that our new inhomogeneous extremal horizons retain a finite entropy density. In fact, the entropy density is always larger for this new IR, than for the extremal RN solutions. Furthermore, due to the boundary conditions at the horizon, the new near horizon geometries still contain an AdS$_2$ factor, however now its radius and transverse space are functions of $x$. We suspect that, unlike AdS$_2 \times \R^2$, these inhomogeneous near horizon geometries are not decoupled scaling solutions on their own, but are partially supported by radial gradients at the horizon.

Finally, we have investigated whether disconnected extremal black holes could nucleate in the interior of the spacetime outside the planar IR horizon. This could occur as the amplitude is increased. Small extremal black holes would nucleate first, and therefore we look for
stable stationary timelike charged geodesics in the inhomogeneous background. With the normalization of (\ref{eq:action}) these would be solutions to
\begin{equation}
U^a\nabla_a U_b = \frac{1}{2}\frac{\mathcal{e}}{\mathcal{m}}F_{ba}U^a\,,
\end{equation}
where $U^a U_a = -1$, and an extremal black hole will have $|\mathcal{e}|=\sqrt{2}|\mathcal{m}|$. We have not found any such geodesics, even at relatively large lattice amplitudes.

\section{\label{sec:ads4}RG stability of AdS$_4$ to periodic deformations}

A natural question raised by the previous sections is whether this nonlinear RG instability exists if there is no net charge, \emph{i.e.} if $\bar{\mu}=0$. The background prior to including a periodic source in this case is AdS$_4$ rather than RN.
This background does not have the exotic emergent $z = \infty$ scaling of AdS$_2 \times \R^2$, and so from the outset we have no reason to expect any interesting low energy physics at finite wavevector. Nonetheless, one might wonder what happens in the limit of large lattice amplitudes. Scale invariance of the underlying microscopic theory now dictates that physical quantities are parametrized by the ratio
\be
\tilde{\alpha}\equiv \frac{\alpha}{k_L} \,.
\ee
Without loss of generality we will fix $k_L = 1$. This system has been studied at finite temperature in \cite{Chesler:2013qla}. Here we will be at exactly $T=0$. The expectation is that as we move into the IR, the solution should globally approach the Poincar\'e horizon of AdS$_4$.

The calculations we perform are similar to those of \cite{Hartnoll:2014cua}, except that here we only consider a single cosine and we work with a lattice in four-dimensional AdS that preserves translational invariance along the field theory direction $w$. As in \cite{Chesler:2013qla}, we can construct solutions perturbatively in $\tilde{\alpha}$, and check agreement with our numerics in the limit where both constructions overlap.

The perturbative construction is best understood in Fefferman-Graham coordinates, in which case the most general line element and gauge field content read:
\begin{equation}
\begin{gathered}
\mathrm{d}s^2 = \frac{L^2}{z^2}\left[-\tilde{A}\mathrm{d}t^2+\tilde{S}_1\mathrm{d}x^2+\tilde{S}_2\mathrm{d}w^2+\mathrm{d}z^2\right] \,,
\\
A = L\,\tilde{P}\,\mathrm{d}t\,,
\end{gathered}
\label{eq:perturbative}
\end{equation}
where $\tilde{A}$, $\tilde{S}_1$, $\tilde{S}_2$ and $\tilde{P}$ are functions of $x$ and $z$ to be determined perturbatively. Since the stress energy tensor in Eq.~(\ref{eqs:motion}) is quadratic in $\tilde{P}$, an expansion about AdS$_4$ in small $\tilde{\alpha}$ takes the form:
\begin{equation}
\begin{gathered}
\tilde{A} = 1+\sum_{j=1}^{+\infty}\tilde{\alpha}^{2i}\tilde{A}^{(2i)}(x,z),\quad \tilde{S}_1 = 1+\sum_{j=1}^{+\infty}\tilde{\alpha}^{2i}\tilde{S}^{(2i)}_1(x,z) \,,
\\
\tilde{S}_2 = 1+\sum_{j=1}^{+\infty}\tilde{\alpha}^{2i}\tilde{S}^{(2i)}_2(x,z)\,,\quad\tilde{P} = \sum_{j=0}^{+\infty}\tilde{\alpha}^{i}\tilde{P}^{(2i+1)}(x,z)\,.
\end{gathered}
\end{equation}

Solving Einstein's equations to third order in $\tilde{\alpha}$ gives\footnote{We went all the way to $10^{\mathrm{th}}$ order in $\tilde{\alpha}$ and found no problem with the perturbative expansion, but the results are too cumbersome to be presented here. In Appendix \ref{app:2} we give the expression for $-\lVert\partial_t\rVert^2/\lVert\partial_w\rVert^2$ at the horizon, the quantity that we will plot in Fig.~\ref{fig:5} below, to tenth order in $A_0$.}
\begin{equation}
\begin{gathered}
\tilde{P}^{(1)}(x,z) = e^{-z}\cos x,
\\
\tilde{A}^{(2)}(x,z) =\frac{1}{8} e^{-2 z} \left(1+2z+2z^2\right)-\frac{1}{8}+\frac{1}{16} e^{-2 z} \left(1+2z+2z^2\right)\cos 2x\,,
\\
\tilde{S}^{(2)}_1(x,z) =\frac{1}{16} e^{-2 z} (1+2z)\cos 2x\,,
\\
\tilde{S}^{(2)}_2(x,z) =\frac{1}{8}-\frac{1}{8} e^{-2 z} \left(1+2z+2z^2\right)+\frac{1}{16} e^{-2 z} \left(1+2z+2z^2\right)\cos 2x\,,
\\
\tilde{P}^{(3)}(x,z) =\frac{1}{512} e^{-3 z} \left(25+44 z+32z^2-25 e^{2 z}\right)\cos x+ \frac{1}{64} e^{-3 z} z (1+z)\cos 3 x\,.
\end{gathered}
\label{eqs:analytic}
\end{equation}
We see no obstruction at any order of perturbation theory in the vicinity of $\tilde{\alpha}=0$ (as we might reasonably have anticipated). We now proceed to corroborate this analytic result with some numerics. Numerically we will find, in addition, no evidence for a phase transition at any finite $\tilde \alpha$, but do find evidence for interesting emergent scaling in the large $\tilde \alpha$ limit.

As discussed in \cite{Hartnoll:2014cua} we need to find a convenient \emph{ansatz} for the De-Turck method. In particular, the so called De-Turck gauge (which is just a generalization of the harmonic gauge) is different from the Fefferman-Graham coordinates of the line element (\ref{eq:perturbative}). We thus consider instead
\begin{equation}
\begin{gathered}
\mathrm{d}s^2 = \frac{L^2}{(1-y)^2}\Big\{(1+y)^2\big[-A\,\mathrm{d}t^2+S_1\,(\mathrm{d}x+F\,\mathrm{d}y)^2+S_2\,\mathrm{d}w^2\big]+\frac{4\,B\,\mathrm{d}y^2}{(1+y)^2}\Big\} \,,
\\
A = L\,P\,\mathrm{d}t\,,
\end{gathered}
\end{equation}
where $y=1$ is the location of the conformal boundary, $y=-1$ is the deep IR, \emph{i.e.} the Poincar\'e horizon, and $A$, $B$, $F$, $S_1$, $S_2$ and $P$ are six unknown functions of $x$ and $y$ to be determined by the numerical scheme. The numerical method we employed here was described in detail in \cite{Hartnoll:2014cua}, so we just quote the final results.

We find good agreement between the perturbative expansion (\ref{eqs:analytic}) and the numerics. For instance, in Fig.~\ref{fig:5} we plot the ratio of minus the norm squared of $\partial_t$ and $\partial_w$ close to the Poincar\'e horizon as a function of $\tilde{\alpha}$ -- the disks correspond to numerical data, and the dashed line corresponds to the analytic prediction (\ref{eqs:analytic}). If this quantity touched zero for some value of $\tilde{\alpha}$ a phase transition or crossover would presumably be triggered. Note that this ratio vanishes on an AdS$_2$ horizon, as it measures the relative `redshift' of time and space. We find no evidence that this occurs. In fact, this quantity seems to scale as $\tilde{\alpha}^{-6}$ for sufficiently large $\tilde{\alpha}$, suggesting an emergent scaling in the limit of large lattice amplitudes.
\begin{figure}[h]
\centering
\includegraphics[height = 0.29\textheight]{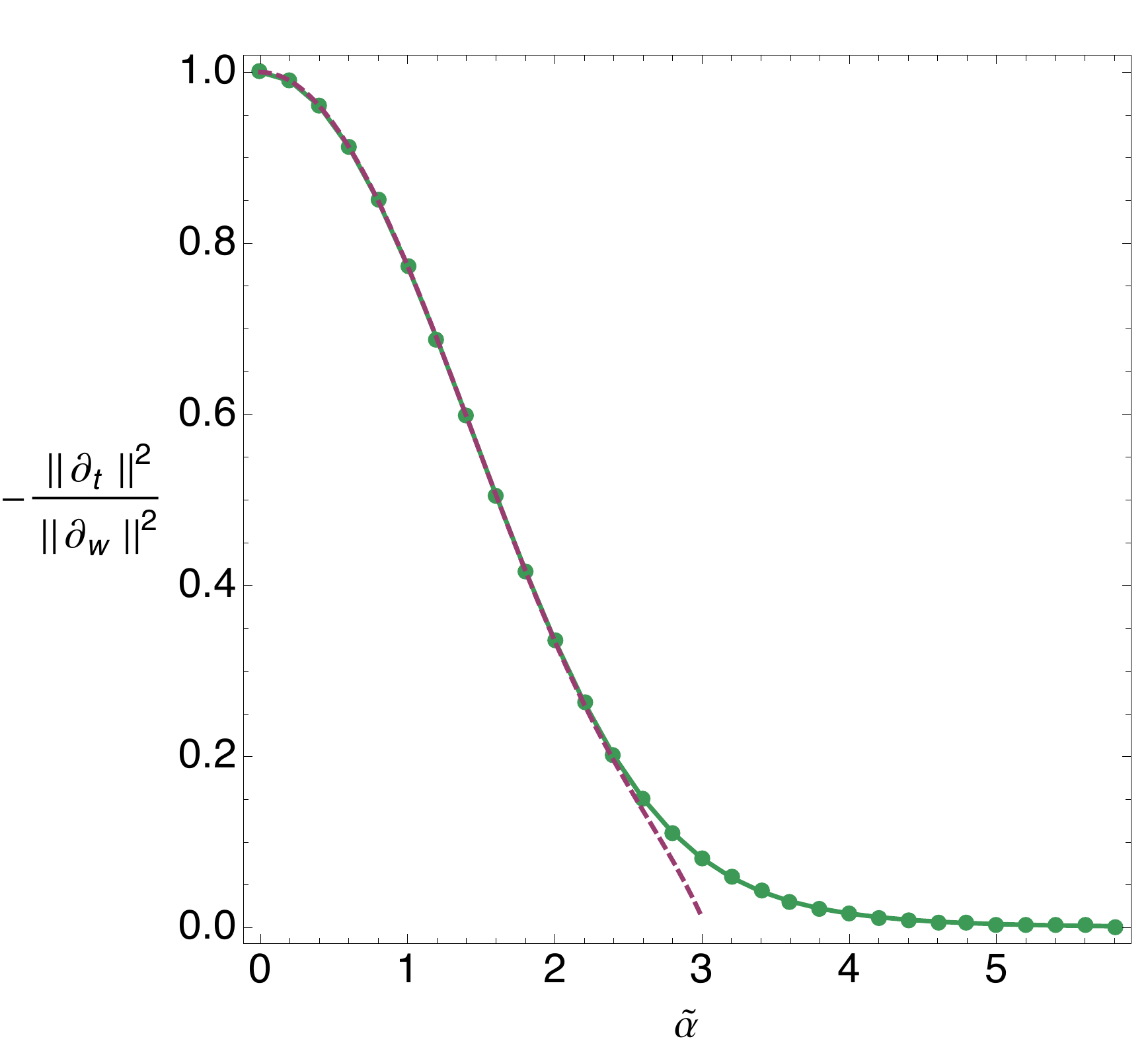}
\caption{\label{fig:5}Ratio of minus the norm squared of $\partial_t$ and $\partial_w$ near the Poincar\'e horizon as a function of $\tilde{\alpha}$. The disks correspond to numerical data, and the dashed line corresponds to the perturbative result (\ref{eqs:analyticapp}), presented in Appendix \ref{app:2} to tenth order.}
\end{figure}
A nonlinear wave never develops at the Poincar\'e horizon, in particular, the IR metric is always that of AdS$_4$, regardless of the values of $\tilde{\alpha}$. Furthermore, as for the case with a net charge, we do not see any stable stationary charged geodesics appearing in the intermediate geometry and therefore disconnected extremal horizons do not form.

The main conclusion of this section is that, as expected, the Poincar\'e horizon of AdS$_4$ is nonlinearly stable against the RG flow generated by a periodic chemical potential.

\section{Discussion: relation to previous and future work}

In this discussion we explain the relationship of our work to previous results. We go on to outline the anticipated condensed matter phenomenology of the inhomogeneous near horizon geometries we have found. We end with comments on future directions.

In moving beyond the simplest extremal horizons, a natural first step is to consider homogeneous but non-translationally invariant horizons \cite{Iizuka:2012iv}. Such solutions capture important aspects of broken translation invariance, in particular a nonzero momentum relaxation rate, while retaining the technical simplification of working with ODEs rather than PDEs. In this homogeneous setup it was shown that if non-translationally invariant couplings are relevant at low energies then they lead to insulating phases, while irrelevant translational invariance breaking leads to metallic phases \cite{Donos:2012js, Donos:2014uba}.

The inhomogeneous results of this paper describe a new IR scenario: a horizon with a tunable amount of inhomogeneity. A finitely inhomogeneous horizon for arbitrarily small and naively irrelevant periodic deformations offers a plausible explanation for the differing behaviors of the conductivity observed in finite temperature numerics on inhomogeneous \cite{Horowitz:2012ky, Horowitz:2012gs} and homogeneous \cite{Donos:2012js, Donos:2013eha} periodic solutions. However, it is possible that nonlinear effects of the kind we have found could arise in the simpler homogeneous cases also.

In terms of inhomogeneous extremal horizons, it seems likely that our two parameter family of solutions is just the tip of an iceberg. As well as having inhomogeneity in both spatial directions, one can imagine superposing an arbitrary number of Fourier modes in the UV. Perhaps it will be possible to find the general inhomogeneous near horizon geometry with a whole function space worth of floppiness.

The survival of a finite low energy inhomogeneity at zero temperature, combined with the persistence of a $z=\infty$ scaling symmetry,  should lead to a finite momentum relaxation rate and hence a nonvanishing zero temperature d.c.~resistivity \cite{Hartnoll:2012rj}. When the IR amplitude of the periodic modulation is small, then the conductivities can be computed by the memory matrix formalism and will lead to a small correction to the results of \cite{Hartnoll:2012rj} at the lowest temperatures: the power law in temperature d.c.~resistivity predicted there will be supplemented by a finite residual resistivity. A residual resistivity due to umklapp (lattice) scattering rather than disorder is exotic, but that is what occurs here.

As the amplitude of the IR lattice grows, the d.c.~resistivity will also grow. The two simplest possibilities are either that the IR amplitude continues to grow without bound, or that there is a phase transition in the IR geometry above some particular amplitude. The former case would correspond to a smooth crossover between metallic and insulating physics as a function of the boundary lattice amplitude. The second case would likely correspond to a (first order?) metal-insulator phase transition. Both cases give a new holographic use of $z=\infty$ scaling, distinct from the directly relevant lattice of \cite{Donos:2012js}, to go between metallic and insulating behavior.

In addition to inhomogeneous periodic sources, modulated bulk modes can also be activated by dynamical instabilities. It is important to distinguish two distinct origins of such instabilities. One possibility is that there are modes in the near horizon AdS$_2 \times \R^2$ geometry that are tachyonic over a range of momentum supported away from zero \cite{Nakamura:2009tf, Donos:2011ff}. These correspond to operators with complex IR scaling exponents. The endpoint of such instabilities is typically the discharging of the extremal horizon, with the new horizon being characterized by some $z < \infty$ \cite{Hartnoll:2011fn}. In these cases the analysis in this paper will not pertain. The second possibility is that the unstable mode is not localized in the near horizon region but rather in the `middle' of the geometry, see e.g. \cite{Dias:2011tj} for a discussion. In this case, condensation does not directly produce a strong backreaction on the extremal horizon. However, inhomogeneous modes will be turned on and will be susceptible to the kind of nonlinear RG flow instability in the near horizon region discussed in this paper.
  
Recent work has constructed disordered horizons by deforming AdS spacetime (at zero charge density) by a marginally relevant disordered boundary coupling \cite{Hartnoll:2014cua}. Disordered couplings can be generated by sums of periodic couplings with random phases. Disorder is able to have a nontrivial effect in the IR even in pure AdS because it involves periodic modes with arbitrarily long wavelength. In this paper we have seen that at a nonzero charge density, a single seemingly irrelevant periodic coupling already has an important effect at low energy. It can therefore be anticipated that disordered boundary couplings in finite density Einstein-Maxwell theory will lead to nontrivial and interesting IR physics.
 
 \section*{Acknowledgements}
 
 It is a pleasure to thank Koushik Balasubramanian, Richard Davison, Diego Hofman, Gary Horowitz and Nabil Iqbal for discussions on related topics and helpful comments on the draft. This work is partially supported by a DOE Early Career Award, by a Sloan fellowship and by the Templeton Foundation.
 
 \appendix
 
 \section{$\varpi$ as a function of $A_0$ for fixed $k_0$}
\label{sec:AA}

\begin{figure}[h]
\centering
\includegraphics[height = 0.29\textheight]{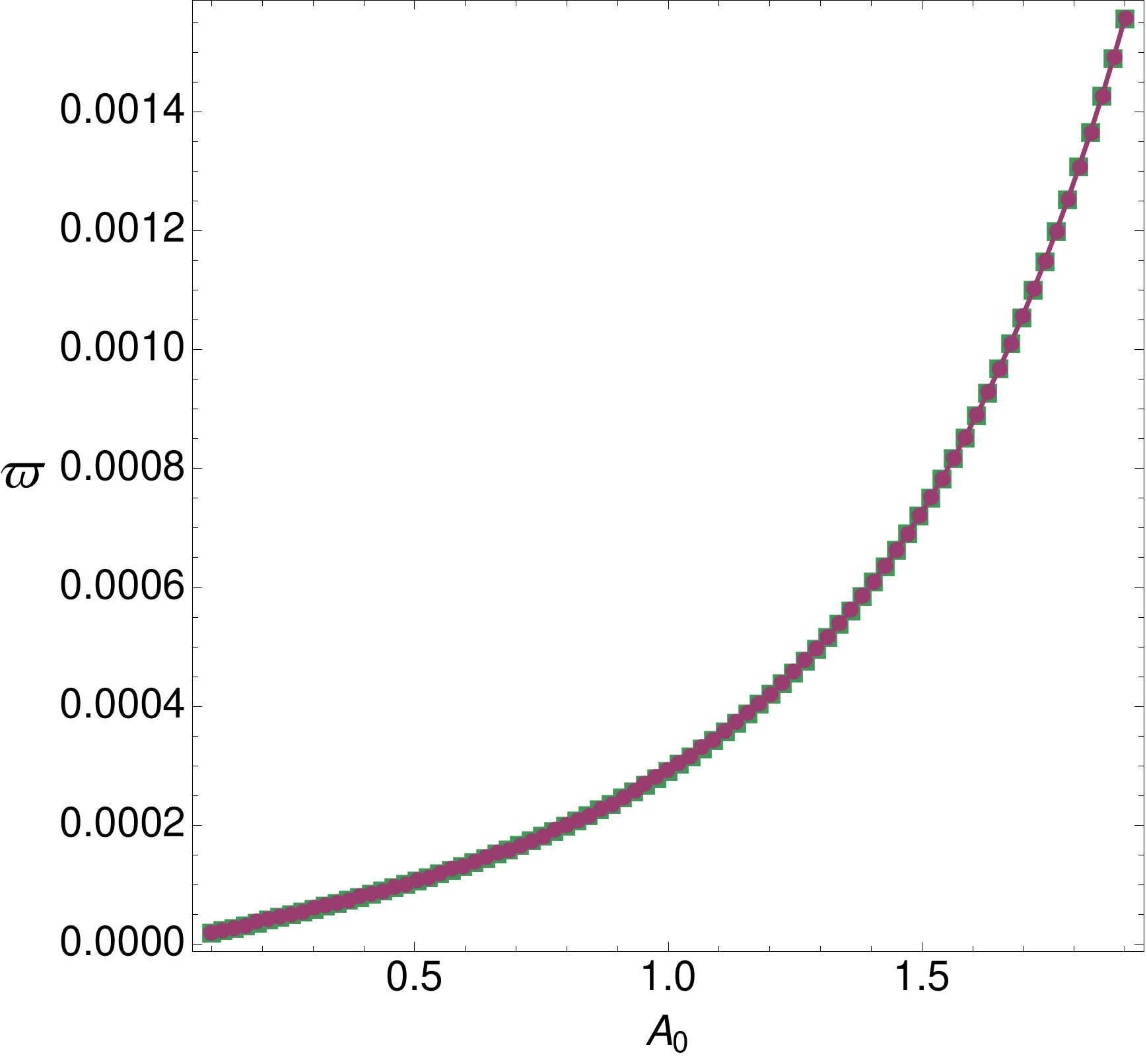}
\caption{\label{apfig:1}$\varpi$ as a function of $A_0$, for fixed $k_0=2$. Circles were computed using a Chebyshev grid with $70\times70$ points, and the squares were computed using a Chebyshev grid with $100\times100$ points.}
\end{figure}

In this appendix we show how $\varpi$ changes as a function of $A_0$, for fixed $k_0=2$. We plot this in Fig.~\ref{apfig:1}. For completeness, we present the results for two different resolutions. The circles have a $70\times70$ Chebyshev grid, whereas squares use a Chebyshev grid with $100\times100$ points.

\section{Ratio of redshifts to 10$^{\mathrm{th}}$ order}
\label{app:2}
\begin{equation}
-\frac{\lVert\partial_t\rVert^2}{\lVert\partial_w\rVert^2}=1-\frac{\tilde{\alpha }^2}{4}+\frac{25 \, \tilde{\alpha }^4}{1024}-\frac{287875 \, \tilde{\alpha }^6}{382205952}-\frac{2697108635 \, \tilde{\alpha}^8}{84537841287168}+\frac{132283354418345243093 \, \tilde{\alpha}^{10}}{182601737180282880000000000} \,.
\label{eqs:analyticapp}
\end{equation}

\end{document}